\pdfoutput=1
\documentclass[12pt, preprint]{aastex}

\def\cc{cm$^{-3}$}
\def\kms{km s$^{-1}$}

\def\h2{H$_2$}
\def\n2h{N$_2$H$^+$}
\def\Ms{M$_\odot$}

\def\13co{$^{13}$CO}
\def\H13CO+{H$^{13}$CO$^+$}
\def\HCO+{HCO$^+$}
\def\c18o{C$^{18}$O}

\def\cm2{cm$^{-2}$}

\def\Mss{$M_{\odot}$\space}
\def\microns{$\mu$m\space}

\def\h2{H$_2$}

\received{2008 September 2}
\begin{document}

\title{  Dichotomy in the Dynamical Status of Massive Cores in Orion}
\author{ T. Velusamy\altaffilmark{1}, R. Peng \altaffilmark{2}, D. Li\altaffilmark{1}, P. F. Goldsmith\altaffilmark{1},
    William D. Langer\altaffilmark{1}    }
%\affil{ Jet Propulsion Laboratory, California Institute of Technology }
\altaffiltext{1}{Jet Propulsion Laboratory, California Institute of
Technology, 4800 Oak Grove Dr.  Pasadena, CA 91109,
Velusamy@jpl.nasa.gov} \altaffiltext{2}{Caltech Submillimeter
Observatory, 111 Nowelo Street, Hilo, HI 96720;
peng@submm.caltech.edu.}

%**********************************************************************************
\begin{abstract}
To study the evolution of high mass cores, we  have searched  for
evidence of collapse motions in a large sample of starless cores in
the Orion molecular cloud.   We used  the Caltech Submillimeter
Observatory telescope to  obtain  spectra of the optically thin
(\H13CO+) and optically thick (\HCO+) high density tracer molecules
in 27 cores with masses $>$ 1 \Ms.  The red- and blue-asymmetries
seen in the line profiles of the optically thick line with respect
to the optically thin line indicate that 2/3 of these cores are not
static. We detect evidence for  infall (inward motions) in 9 cores
and    outward motions for 10 cores, suggesting a dichotomy in the
  kinematic state of the non-static cores in this sample.     Our results provide an
important observational constraint on the   fraction of collapsing
(inward motions) versus non-collapsing (re-expanding) cores for
comparison with model simulations.

\end{abstract}
\setcounter{footnote}{0}

%*******************************************************************************
\section{INTRODUCTION}

 In giant molecular clouds, the formation of massive cold   molecular
 cores   can be induced by gravoturbulent cloud fragmentation (c.f. Mac Low \& Klessen 2004).
    McKee \& Tan (2003) envisage that these
cores are molecular condensations in a turbulence-supported
quasi-equilibrium state,  which ultimately form a single protostar
or a gravitationally bound cluster of massive protostars.   This
important stage in star formation is not well understood due to lack
of observational data on prestellar cores on the verge of collapse.
To study this stage we obtained new data on the kinematic state for
a sizable number of Orion cores.    Li et al. (2007, Paper II)
identified 51 dust cores in a 350 \microns submillimeter continuum
survey of the quiescent regions of the Orion molecular cloud using
the SHARC II camera at the Caltech Submillimeter Observatory (CSO).
A combination of the enhanced spatial resolution analysis of the
submm data using the HiRes deconvolution tool (see  Velusamy et al.
2008) and our knowledge of the temperature from the ammonia mapping
by Li et al. (2003, Paper I) enabled us to determine relatively
accurately the key physical parameters  (including size  and mass)
of cores in this representative sample of starless cores in a high
mass star forming cloud. This Orion core sample is a collection of
resolved or nearly resolved cores, with a mean mass of 9.8
M$_\odot$, which is 1 order of magnitude higher than that of
resolved cores in low-mass star-forming regions. The majority of the
cores appear to be too massive to be thermally supported as
Bonner-Ebert spheres for T$_K$ $<$ 30K (Paper II), or even with
turbulence $\sim$ 1.5 \kms\space (for reference the mean linewidth
observed in these cores is 0.78 \kms). Therefore, magnetic support
may be important in these cores (c.f. Mouschovias, 1991).  Here we
present new spectroscopic observations to determine the kinematics
of these cores. We discuss the prevalence of dynamically unstable
high mass cores in this sample and the equally large proportion of
cores showing either collapse or expansion.

\section{Observations }

  We selected a
subset of  27 cores  from the sample in Paper II which have mass $>$
1 \Ms. The observations were made at the 10.4 m telescope of the
Caltech Submillimeter Observatory (CSO) at Mauna Kea, Hawaii, in
March and December, 2007.  The cores and their   coordinates were
selected from Table 1 in Paper II.  We observed both \HCO+(3-2) at
267.557620 GHz, and \H13CO+(3-2) at 260.255478 GHz  in all 27 cores.
We used the  230 GHz band heterodyne receiver and
 the 1024 channel  acousto-optic spectrometer with a
bandwidth of 50 MHz.  The FWHM antenna beam   was 26$\arcsec$ and
the main beam efficiency was 0.66. To improve the SNR on the
optically thin \H13CO+(3-2) spectra, 3 adjacent velocity channels
were coadded.    The velocity $V_0$ and line width $\Delta V_{thin}$
for each core were obtained using Gaussian profile fit to the
optically thin \H13CO+(3-2) lines.   The spectra of all 27 cores are
shown in     Figure 1.  The spectra of the cores are presented in
three groups as a function of their asymmetry (see Section 3) in the
\HCO+ line with: (a) blue shifted peak emission, (b)  red shifted
peak emission, and (c) no significant asymmetry with respect to the
\H13CO+   spectrum. The spectra within each group are arranged in
order of their masses.

For most of the cores the line widths for \H13CO+ are $>$ 0.5
\kms\space and appear to be dominated by turbulence compared to
\HCO+ thermal line widths $<$ 0.2 \kms\space for T$_K$ $<$ 20 K.
Thus the majority of our cores are supersonic.     Using the total
linewidth as an    upper limit  of the equivalent sound speed we
estimated approximate   Bonnor-Ebert (B-E) critical mass (M$_{B-E}$)
for each core (see Paper II).  Only 3 cores have masses within a
factor 2.5 of their critical masses while the other cores all have
masses well above their critical masses for stability.

\section{Results and Discussion}

To study the kinematic status of the  cores     we use the spectral
line asymmetry as observed between a pair of optically thick and
thin molecular transitions.  We define the red- and blue- asymmetry
by the peak emission in the optically thick line appearing red-ward
and blue-ward, respectively, with respect to the peak in the
optically thin spectrum.  The method of spectral line asymmetry
(Leung \& Brown 1977; Zhou et al. 1992) has commonly been  used to
identify collapse candidates. Under certain assumed collapse
conditions (e.g. Shu et al. 1987) an optically thick line  (e.g.
\HCO+(3-2)) in a collapsing core shows a blue-asymmetry in the line
profile observed toward the collapse center, while an optically thin
line (e.g. \H13CO+(3-2)) appears symmetric.   Gregersen et al.
(1997) have used the same pair of lines to search for infall in 19
low mass cores.  They detected blue-asymmetry   in 9 Class 0
objects, and red-asymmetry in 2 objects. Using the   HCO$^+$(3-2)
line as the optically thick tracer, in their search for infall
candidates in low mass cores, Mardones et al. (1997) detected
blue-asymmetry in 8, red asymmetry in 3, and found  one (VLA 1623)
to be inconclusive. Fuller et al. (2005) observed  77 submillimeter
sources associated with a sample of IRAS selected objects, all
believed to be high mass protostellar objects. They  identified 22
infall candidates none of which show any red-asymmetry.
    Thus red-asymmetries
are rare in low mass cores (c.f. Lee et al. 2001) as well as in high
mass  cores containing  protostars and the few cores showing red
asymmetries are caused by outflows (Gregersen et al. 1997).
 In contrast,   in our   Orion sample of high mass starless cores
(not containing any known protostars)   approximately 2/3 show blue-
or red-asymmetries, roughly in equal numbers.

In principle, information on the velocity structure v(r) within the
core  is contained in   the line profile averaged over the core. In
the spectra shown in Figure 1, the line-of-sight  motions within the
core produce Doppler shifts and features in the line profiles
averaged over the core. A static spherical core  displays a
symmetric spectral line profile for low optical depths.  At large
optical depths it exhibits a self-absorption dip at the center with
symmetric blue and red peaks. Radial variations in physical
conditions, e.g. temperature, density, molecular abundances, line
widths, do not affect the symmetry of the observed spectrum relative
to the central velocity of the core.  By looking at whether the line
emission is stronger in the red or blue, one can distinguish between
infall and expansion in the core (Leung \& Brown, 1977; Zhou et al.
1993). Schematically, in the case of radial infall  the red and blue
emissions come from, respectively, the hemispheres closer to and
farther from the observer. If the excitation temperature (T$_{ex}$)
decreases radially outward from the core center, as  expected
  for \HCO+(3-2), the line-of-sight from the observer at red
shifted frequencies first encounters cooler gas (at greater distance
from the core center), while at blue shifted frequencies warmer gas
(closer to the core center) is encountered first.   Therefore,
subject to the caveat on the velocity field discussed below, the
blue emission is stronger than the red,  producing a blue-asymmetry.
In the case of radial expansion of the core the blue and red
hemispheres are reversed and the red emission is stronger, resulting
in a red-asymmetry.

The effects of the self-absorption on the red or blue peaks of the
spectrum depends critically on the velocity field as a function of
radius. There is no self-absorption in the peaks if v(r) $\propto$
r, as every line-of-sight velocity component corresponds to a single
point along the line-of-sight.    However, for a velocity field
having a power law with negative index, for example, v(r) $\propto$
r$^{-0.5}$, as in the case of inside-out collapse model (Shu et al.
1987), every line-of-sight velocity component corresponds to two
points at different distances from the core center (see Fig. 8 in
Zhou et al. 1993). Thus this velocity law favors a stronger
self-absorption in the red as the   two iso-velocity points in the
red hemisphere correspond to a high and low T$_{ex}$, respectively,
farther and closer to the observer (see above). To produce the red-
and blue-asymmetries by self-absorption, we require
   a favorable radial variation of not only velocity, v(r) but also
  excitation temperature, T$_{ex}$(r).  In a region where the HCO$^+$ emission is produced,
 the excitation of \HCO+ (3-2) is subthermal and is dominated by the radial
 increase in density towards the core center. Thus T$_{ex}$(r) will
always increase radially toward the center essentially independent
of the temperature profile (Paper I).   LVG model calculations show
that such conditions in a family of model clouds with density $\sim$
10$^5$ - 10$^6$ \cc\space and  T$_K$ $\sim$ 15 to 25 K,  can produce
an infall blue-asymmetry. We can rule out rotation as the cause of
asymmetry in these cores as they are spatially unresolved and
spectra were obtained toward the center of rotation of the cores
(Menten et al. 1987, Adelson \& Leung 1988).

Clearly, the asymmetries observed in the spectra in Figure 1 show
that a majority of these cores (19 out of   27) are not static but
have inward or outward motions. The interpretation of the
blue-asymmetry as infall is well understood and has been modeled
extensively (e.g. Choi et al. 1995). The interpretation of the
red-asymmetry   as outward expansion is less obvious. We can rule
out  interpreting red-asymmetry as   infall because it requires
abnormal  conditions in which the excitation temperature will  be
lower near the core center.  On the other hand, a reasonable
interpretation of red-asymmetry is outward (expansion) motion in the
core with a velocity law v(r) $\propto$ r$^{-\beta}$ where $\beta$
$>$ 0.
 Though a detailed interpretation of the dynamics associated with asymmetries
in the line profiles   will be model dependent, considering the
above scenarios,    here we assume that the cores showing red- or
blue-asymmetries have either inward or outward motions. Among the
   27 cores for which we have both the optically thick and
thin spectral line profiles, 9 show blue-asymmetry,  10 show
red-asymmetry, while the remaining 8 have  roughly symmetrical
profiles with a single velocity peak. Following  Mardones et al.
(1997) we  quantify the red- and blue- asymmetry by the asymmetry
parameter:
\begin{equation}
\delta V  = (V_{thick} - V_{thin})/\Delta V_{thin}
\end{equation}
where $V_{thick}$ is the velocity of the peak of the opaque
(HCO$^+$)
 line, $V_{thin}$ is the velocity of the peak of the optically thin
(\H13CO+)   line, and $\Delta V_{thin}$ is the line width of the
thin line; $\delta V$ is indicated on each spectra in Figure 1.  In
Figure 2(a) we show the distribution of the observed asymmetry
parameters. The distribution is bimodal with peaks having distinct
blue- ($\delta V$ negative) and red- ($\delta V$ positive)
asymmetries.

A fairly clear demarcation between the red- and blue- asymmetries as
a function of core mass is evident in Figure 2b.  The higher mass
cores (with the exception of ORI1\_13 and ORI2\_2) tend to show a
red-asymmetry while the lower mass cores show a  blue-asymmetry. If
we interpret the red-asymmetry  as outward motion (expansion) within
the core it is conceivable that these cores  will fragment and
eventually form only multiple lower mass stars. At the high end of
the mass distribution,     all of the 12 cores, having masses $>$ 15
\Ms, show either red- or blue- asymmetry and are
 either collapsing (3 cores) or expanding (9 cores).    Thus for the high mass
 cores ($>$ 15\Ms) we observe a larger proportion
(by a factor 3) of cores showing expansion compared to those showing
collapse.  In contrast, among the 16 lower mass cores ($<$ 15\Ms)
only one (ORI1\_17)  shows evidence for expansion, while 6 cores
show collapse and 8 are static. Assuming the cores showing
blue-asymmetry will eventually form stars (3/12 for $>$ 15\Mss and
6/16 for $<$ 15\Ms), we can estimate that 1/3 of the cores ($\sim$
32\% of the total mass in our present sample of 27 cores in Orion)
will form high mass stars.

   Our Orion sample represents a population of   massive molecular
cloud   cores   in which the majority of the cores are dynamic
 out-of-equilibrium structures. We detect cores in roughly
equal proportion showing  all three kinematic states:   static,
  outward, and  inward
motion. Therefore,   we may be seeing, statistically,  a snapshot of
a population of cores exhibiting oscillatory motion. Oscillations in
cores about a stable equilibrium configuration, producing the red-
and blue-asymmetries, have been observed in low mass cores (Redman
et al. 2006; Keto et al. 2006; Broderick et al. 2007). The masses of
our cores, with the exception of a few,   far exceed the critical
masses for equilibrium as isothermal Bonnor-Ebert spheres (see Paper
II, Fig.8). However,  it is conceivable that  they are magnetically
supported. Although they are thermally supercritical, they could
still be magnetically subcritical having equilibrium configurations
(c.f. Mouschovias, 1991). Under such conditions they are easily
excited by the external interstellar medium to exhibit oscillatory
behavior, consistent with the distribution of the kinetic state of
cores observed in our sample.  The majority of cores above the
median mass of our sample have red asymmetric profiles indicative of
expansion.   Numerical models also suggest that massive molecular
cloud cores are in general  likely to be dynamic, out-of-equilibrium
structures, rather than quasi-hydro/magneto-static structures
(V\'{a}zquez-Semadeni et al. 2005). In their models for the
evolution of clumps and cores formed as turbulent density
fluctuations, the dynamically compressed regions must either proceed
to collapse right away or re-expand. These models predict that not
all cores observed in molecular clouds will necessarily form stars
and that a class of ''failed cores'' should exist, which will
eventually re-disperse. Thus we should  see some fraction of   cores
displaying outward as well as inward motions. Our results do show a
larger detection of kinematic signatures with inward and outward
motions than in previous studies, particularly in low mass starless
cores (Lee et al. 2001). This result may not be unique to  Orion and
in general may be more common in massive  than in low mass cores,
but we are better able to detect kinematic signatures in Orion due
to its proximity. Our results provide an important observational
constraint on the   fraction of collapsing (inward motions) versus
non-collapsing (re-expanding) cores for comparison with model
simulations.

 \acknowledgments
We thank Dr. Konstantinos Tassis for comments on the dynamics of
magnetically supported cores.  We thank Dr. Jorge Pineda for running
cloud models incorporating specific large-scale velocity fields to
confirm the expected behavior of the red- and blue-asymmetries in
the Orion cores.  The research described in this paper was carried
out at the Jet Propulsion Laboratory, California Institute of
Technology, under a contract with the National Aeronautics and Space
Administration and supported by a grant from NASA Astronomy and
Physics Research and Analysis Program. Research at the Caltech
Submillimeter Observatory is supported by NSF grant AST-0229008.

\clearpage

%*********************************************************************************
%Figures
%*********************************************************************************
%Fig 1
\begin{figure}          \label{fig1}
    \includegraphics[scale=0.7]{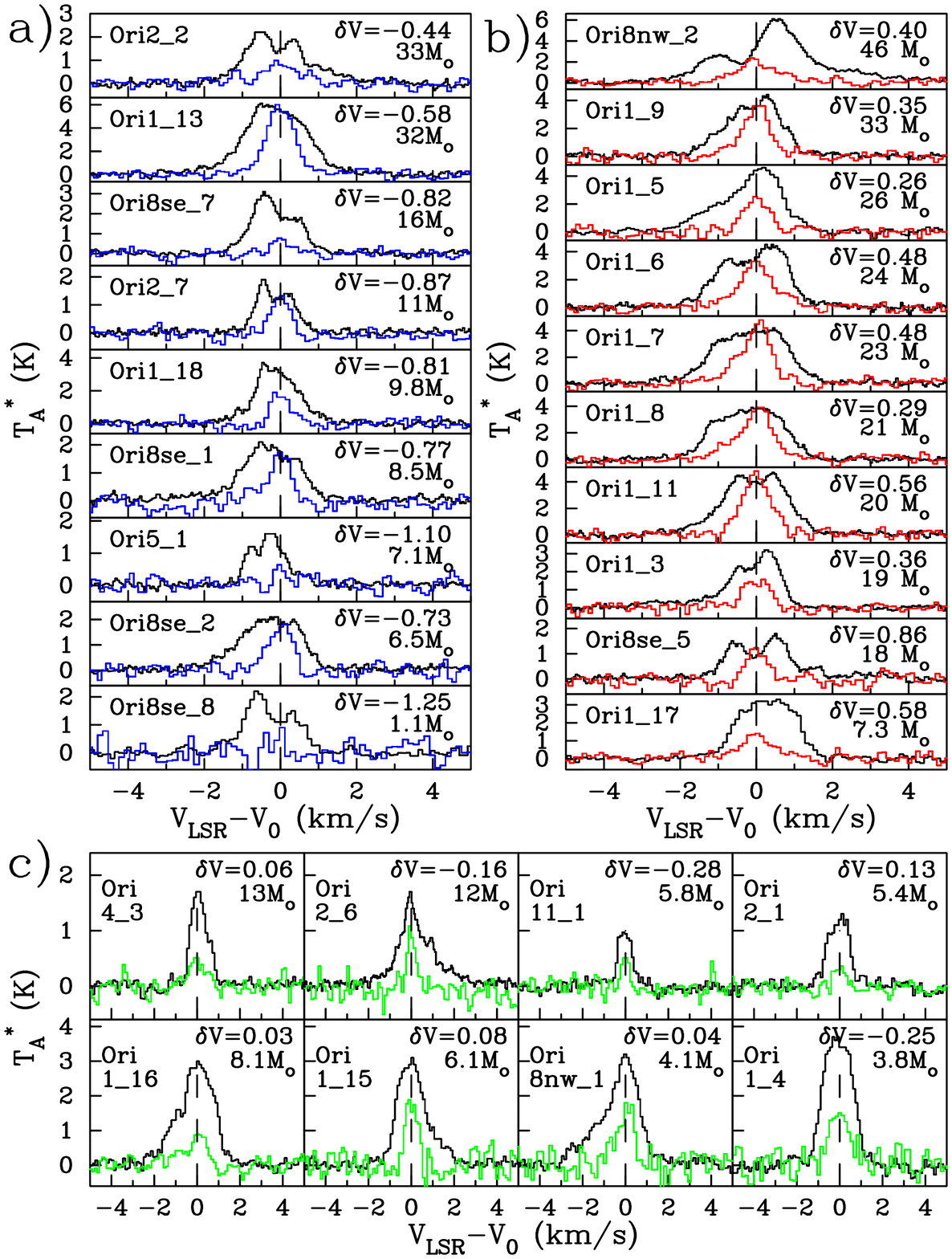}
    \caption{The  spectra of  the optically thick, HCO$^+$(3-2)    and thin, \H13CO+(3-2)   lines
    for 27 cores     in our sample with mass $>$ 1M$_\odot$.  The     HCO$^+$     spectra are shown in black and the
     \H13CO+      spectra in color. The    \H13CO+ intensities  have been multiplied by a factor of 3.
     The channel velocities in each panel    are $ V_{LSR} - V_0$ where
     $V_0$      is    the peak velocity  of  \H13CO+ profile. The mass of the core,
     and the asymmetry parameter $\delta V$      (see Eq. 1)    are indicated.
     The cores are identified as given in Paper II. The spectra are grouped into cores with: (a)   blue-asymmetry;
     (b) red-asymmetry; (c) no significant
    asymmetry.
        }
\end{figure}

\begin{figure}[htp]         \label{fig2}
\includegraphics[ scale=0.75]{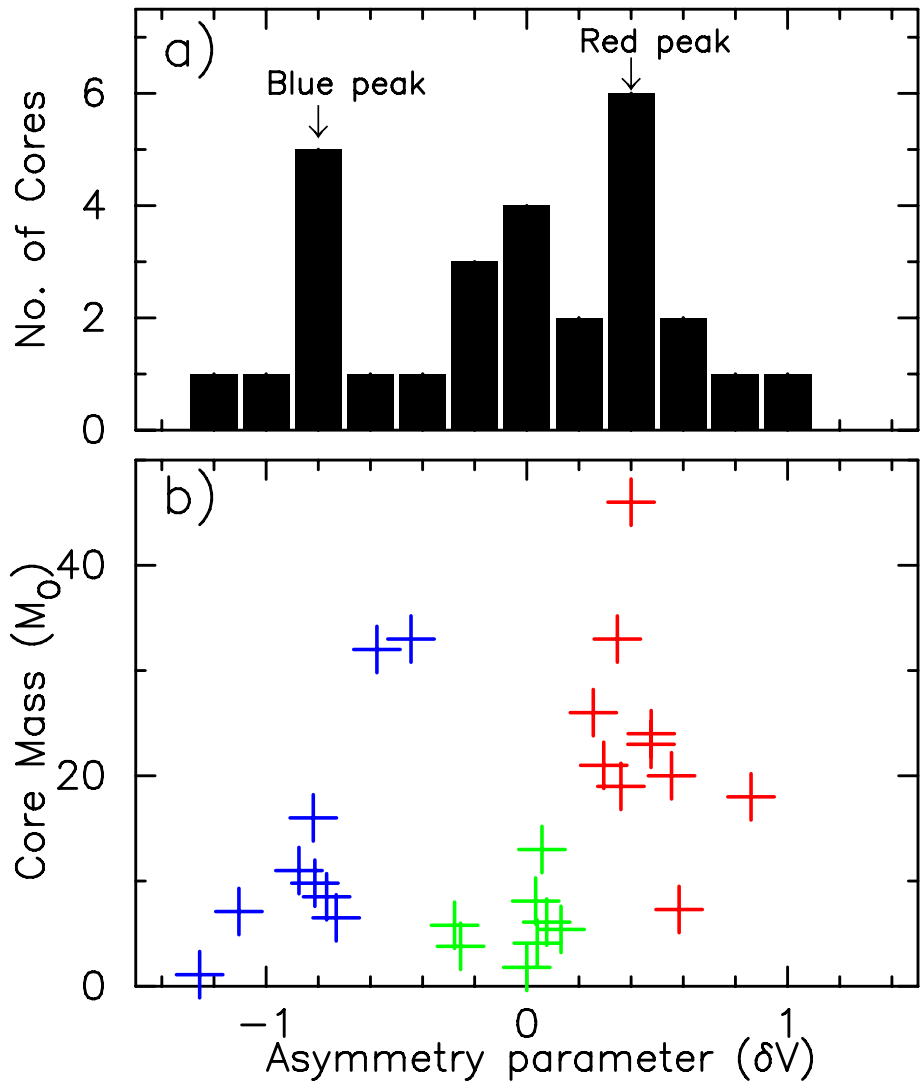}
\caption{(a) The   distribution of the  velocity asymmetry parameter
$\delta V$  defined in eq. (1)  for the full sample of 27 cores
obtained using   optically thick  (HCO$^+$) and thin (\H13CO+)
lines. Note the dichotomy in the kinematic state of the non-static
high mass cores (b) Asymmetry parameter {\it versus} core mass. The
blue and red crosses correspond to cores showing   blue- and
red-peak asymmetry, respectively in  (HCO$^+$) with respect to
(\H13CO+) as shown Figs 1(a), and 1(b).   The green crosses
represent cores (Fig. 1(c)) that do not show  any  significant
asymmetry.    Note  large proportion of the high mass cores have a
red-asymmetry. }
\end{figure}

\end{document}